\documentstyle[seceq]{ptptex}

\title{
Particle Methods in Astrophysical Fluid Dynamics
}

\author{
Frederic A.~{\sc Rasio}
}

\inst{
Department of Physics, MIT, Cambridge, MA 02139, USA
}

\recdate{
November 19, 1999
}

\abst{
Particle methods play an important role in the study of a wide variety 
of astrophysical fluid dynamics problems. The different methods 
currently in use are all variants of the 
so-called Smoothed Particle Hydrodynamics (SPH) scheme
introduced by Lucy and Gingold \& Monaghan more than twenty years ago.
This paper presents a complete introduction to SPH in its modern
form, and discusses some of the main numerical properties of the scheme.
In particular, the convergence properties of SPH are studied,
as a function of the number of particles $N$ and the number of 
interacting neighbors $N_N$, using a simple analysis based on sound waves.
It is shown that consistency of SPH (i.e., convergence towards a
physical solution) requires both $N\rightarrow\infty$
and $N_N\rightarrow\infty$, with the smoothing length 
$h\rightarrow 0$, i.e., $N_N/N\rightarrow 0$. 
}

\begin{document}

\maketitle

\section{Introduction}

Many important problems in modern astrophysics involve fluids moving freely in
3D under the influence of self-gravity and pressure forces. These
problems are best approached numerically using a Lagrangian formulation
where the fluid system is represented by a large number of particles.
The most popular scheme, known as Smoothed Particle Hydrodynamics (SPH), 
is presented in this paper.
The key idea of SPH is to calculate pressure gradient forces by kernel
estimation, directly from the particle positions, rather than by finite
differencing on a grid.
The basic form of SPH was introduced more than twenty years ago by 
Lucy (1977) and Gingold \& Monaghan (1977), who used it to study dynamical 
fission instabilities in rapidly rotating stars. Since
then, a wide variety of astrophysical fluid dynamics processes have been
studied numerically in 3D using SPH (see Monaghan 1992 for an overview). 
These include many stellar interaction processes such as 
binary star coalescence (e.g., Rasio \& Shapiro 1994, 1995; Rasio \& Livio 1996;
Zhuge et al.\ 1996; Rosswog et al.\ 1999) 
and stellar collisions 
(e.g., Lai, Rasio, \& Shapiro 1993; Lombardi, Rasio, \& Shapiro 1996; 
Bailey \& Davies 1999), as well as
star formation and planet formation (e.g., Nelson et al.\ 1998; Burkert et al.\ 1997), 
supernova explosions (e.g., Herant \& Benz 1992; Garcia-Senz et al.\ 1998), 
large-scale cosmological structure formation (e.g., Katz et al.\ 1996;
Shapiro et al.\ 1996), and galaxy formation (e.g., Katz 1992; Steinmetz 1996).

Because of its Lagrangian nature,
SPH presents some clear advantages over more traditional 
grid-based Eulerian methods for calculations of astrophysical fluid flows. 
Most importantly, fluid advection, even for objects with a sharply defined surface
such as stars, is accomplished without difficulty in SPH, since the
particles simply follow their trajectories in the flow. In contrast, 
to track accurately, for example, the orbital motion of two stars across a 
large 3D grid, can be quite tricky, and the stellar surfaces then require a special
treatment (to avoid ``bleeding''). SPH is
also very computationally efficient, since it concentrates the
numerical elements (particles) where the fluid is at all times,
not wasting any resources on emty regions of space. 
This is particularly important for processes involving a large
dynamic range in densities, such as gravitational collapse and
fragmentation.
For this reason, with given computational resources, SPH provides higher
averaged spatial resolution than grid-based calculations, although
Godunov-type schemes such as PPM (the Piecewise-Parabolic Method; see,
e.g., Woodward 1986) typically provide better 
resolution of shock fronts.
SPH also makes it easy to track the hydrodynamic ejection of matter to 
large distances from central dense regions. Sophisticated nested-grid 
algorithms are often necessary to accomplish the same with grid-based
methods (see, e.g., Ruffert 1993). 
Finally, SPH makes it easy to track the evolution of any
passively advected scalar quantity, such as the chemical composition of
the fluid (see, e.g., Lombardi et al.\ 1995, 1996, for an application to
the study of hydrodynamic mixing during stellar collisions).

\section{Basic Equations and Properties of the SPH Scheme}

\subsection{SPH from a Variational Principle}

A straightforward derivation of the basic SPH equations
can be obtained from a Lagrangian formulation of hydrodynamics
(Gingold \& Monaghan 1982). Consider for
simplicity the adiabatic evolution of an ideal fluid with equation of state
\begin{equation}
p=A\rho^\gamma, \label{eos}
\end{equation}
where $p$ is the pressure, $\rho$ is the density, $\gamma$ is the adiabatic
exponent, and $A$ (assumed here
to be constant in space and time) is related to the specific
entropy ($s \propto \ln A$). The Euler equations of motion,
\begin{equation}
{d\vec{v}\over dt}=
   {\partial\vec{v}\over\partial t}+(\vec{v}\cdot\nabla)\vec{v}
     =-{1\over\rho}\nabla p,
\end{equation}
can be derived from a variational principle with the Lagrangian
\begin{equation}
L=\int\left\{{1\over2}v^2 - u[\rho(\vec{r})]\right\}\,\rho\,d^3x.
\end{equation}
Here 
$u[\rho]=p/[(\gamma-1)\rho]=A\rho^{\gamma-1}/(\gamma-1)$ is the specific internal 
energy of the fluid. 

The basic idea in SPH is to use the discrete representation 
\begin{equation}
L_{SPH}=\sum_{i=1}^N\, m_i\left[{1\over2}v_i^2 -u(\rho_i)\right] \label{lsph}
\end{equation}
for the Lagrangian, where the sum is over a large but discrete number of small
fluid elements, or ``particles,'' covering the volume of the fluid. Here $m_i$ 
is the mass and $\vec{v}_i$ is the velocity of the particle with position 
$\vec{r}_i$.  For expression~(\ref{lsph}) to become the Lagrangian of a system 
with a finite number $N$ of
degrees of freedom, we need a prescription to compute the density 
$\rho_i$ at the
position of any given particle $i$, as a function of the masses and positions of
neighboring particles.

In SPH, the density at any position is calculated as the
local average
\begin{equation}
\rho(\vec{r})=\sum_j m_j W(\vec{r}-\vec{r}_j;\, h),\label{rho}
\end{equation}
where $W(\vec{r};\,h)$ is a smoothing kernel of
width $\sim h$.  Necessary constraints on the kernel
$W(\vec{r};\,h)$ are that (i) it integrates to unity (consequently the
integral of eq.~(\ref{rho}) over all space automatically gives the
total conserved mass of the system), and (ii) it approaches the Dirac delta
function $\delta(\vec{r})$ in the limit where $h\rightarrow 0$.
In practice, smoothing kernels with finite supports are almost always
used, so that a finite number $N_N$ of particles around $\vec{r}$ 
contribute to the estimate of $\rho(\vec{r})$.

Eq.~(\ref{rho}) gives, in particular, the density in the vicinity of 
particle $i$ as $\rho_i=\rho(\vec{r}_i)$, and we can now obtain the equations of motion
for all the particles. Deriving the Euler-Lagrange equations from $L_{SPH}$ we get
\begin{equation}
{d\vec{v}_i\over dt}=-\sum_j m_j \,\left({p_i\over\rho_i^2}
                   +{p_j\over\rho_j^2}\right)\, \nabla_i W_{ij}, \label{simple}
\end{equation}
where $W_{ij}= W(\vec{r}_i-\vec{r}_j;\, h)$ and we have assumed that
the form of $W$ is such that $W_{ij}=W_{ji}$.
The expression on the
right-hand side of eq.~(\ref{simple}) is a sum over neighboring particles (within a 
distance $\sim h$ of $\vec{r}_i$) 
representing a discrete approximation to the pressure gradient
acceleration $[-(1/\rho)\nabla p]_i$ for particle $i$.

The following energy and momentum conservation laws are satisfied {\em exactly\/}
by these simple SPH equations of motion:
\begin{equation}
{d\over dt}\left(\sum_{i=1}^N m_i \vec{v}_i \right) =0,
\end{equation}
and
\begin{equation}
{d\over dt}\left(\sum_{i=1}^N m_i\, [{1\over2}v_i^2 +u_i]\right) =0,
\end{equation}
where $u_i=p_i/[(\gamma-1)\rho_i]$.
Note that energy and momentum conservation in this simple version of
SPH is independent of the number of particles $N$.

Typically, a full implementation of SPH for astrophysical problems will add
to eq.~(\ref{simple}) a treatment of self-gravity (e.g., using one of the many 
grid-based or tree-based algorithms developed for $N$-body simulations) and an
artificial viscosity term to allow for entropy production in shocks.  
In addition, we have assumed here that the smoothing length
$h$ is constant in time and the same for all particles. In practice, individual
and time-varying smoothing lengths $h_i(t)$ are almost always used, so that the
local spatial resolution can be adapted to the (time-varying) density of SPH particles
(see Nelson \& Papaloizou 1994 for a rigorous derivation of the equations of 
motion from a variational principle in this case). Other derivations of the SPH
equations, based on the application of smoothing operators to the fluid equations
(and without the use of a variational principle), are also possible (Monaghan 1985;
Hernquist \& Katz 1989; Monaghan 1992).

\subsection{Basic SPH Equations}

In this section, we summarize the basic equations for various forms
of the SPH scheme currently in use,
incorporating gravity, artificial viscosity, and individual smoothing lengths.

\subsubsection{Density and Pressure}

The SPH estimate of the fluid density at $\vec{r}_i$ is
calculated as $\rho_i=\sum_j m_j W_{ij}$ [cf.\ eq.~(\ref{rho})].  
Many recent implementations of
SPH use a form for $W_{ij}$ proposed by Hernquist \& Katz (1989),
\begin{equation}
W_{ij}={1\over2}\left[W(|\vec{r}_i-\vec{r}_j|;\,h_i)+W(|\vec{r}_i-
\vec{r}_j|;\,h_j)\right].
\end{equation}
This choice guarantees symmetric weights $W_{ij}=W_{ji}$ even between
particles $i$ and $j$ with different smoothing lengths.
For the smoothing kernel $W(r;\,h)$, the cubic spline
\begin{equation}
W(r;\,h)={1\over\pi h^3}
\cases{1-{3\over2}\left({r\over h}\right)^2
      +{3\over4}\left({r\over h}\right)^3,        & $0\le{r\over h}<1$,\cr 
{1\over4}\left[2-\left({r\over h}\right)\right]^3,& $1\le{r\over h}<2$,\cr
      0,                                          & ${r\over h}\ge2$,\cr} \label{WML}
\end{equation}
introduced by Monaghan \& Lattanzio (1985) is a common choice.
Eq.~(\ref{WML}) is sometimes called a ``second-order accurate'' kernel.
Indeed, when the true density $\rho(\vec{r})$ of the fluid is represented by an
appropriate distribution of particle positions, masses, and smoothing
lengths, one can show that $\rho_i=\rho(\vec{r}_i)+O(h_i^2)$
(see, e.g., Monaghan 1985). Spherically symmetric kernels such as
that of eq.~(\ref{WML}) can lead to loss of spatial resolution for
highly anisotropic flows (as in, e.g., cosmological pancake-type collapse). 
Adaptive, anisotropic kernels can be used 
for those problems (Fulbright et al.\ 1995; Shapiro et al.\ 1996; Owen et al.\ 1998).

Depending on which thermodynamic evolution equation is integrated (see
\S2.2.4 below), particle $i$ also
carries either the parameter $u_i$, the internal energy per
unit mass in the fluid at $\vec{r}_i$, or $A_i$, the entropic variable,
a function of the specific entropy in the fluid at $\vec{r}_i$.
Although arbitrary
equations of state can be implemented in SPH, here, for simplicity,
we consider only polytropic equations of state:
the pressure $p_i$ at $\vec{r}_i$ is related to the density by
\begin{equation}
p_i=(\gamma-1)\,\rho_i\, u_i,
\end{equation}
or
\begin{equation}
p_i=A_i\,\rho_i^\gamma.
\end{equation}
The speed of sound in the fluid at $\vec{r}_i$ is $c_i=(\gamma
p_i/\rho_i)^{1/2}$.  

\subsubsection{Dynamical Equations and Gravity}

Particle positions are updated either by
\begin{equation}
{d\vec{r}_i \over dt}= \vec{v}_i, \label{rdot}
\end{equation}
or the more general XSPH method
\begin{equation}
{d\vec{r}_i\over dt} = \vec{v}_i+\epsilon \sum_j m_j{\vec{v}_j-\vec{v}_i\over
\rho_{ij}}W_{ij} \label{XSPH}
\end{equation}
where $\rho_{ij}=(\rho_i+\rho_j)/2$ and $\epsilon$ is a constant
parameter in the range $0 < \epsilon < 1$ (Monaghan 1989).  Eq.~(\ref{XSPH}), 
in contrast to eq.~(\ref{rdot}), changes particle
positions at a rate closer to the local smoothed velocity.  The XSPH
method was originally proposed as a way to minimize spurious interparticle
penetration across the interface of two colliding fluid streams.

Generalizing eq.~(\ref{simple}) to account for gravitational forces 
and artificial viscosity (hereafter AV), 
the velocity of particle $i$ is updated according to
\begin{equation}
           {d\vec{v}_i\over dt} = \vec{a}^{(Grav)}_i+\vec{a}^{(SPH)}_i
\end{equation}
where $\vec{a}^{(Grav)}_i$ is the gravitational acceleration and
\begin{equation}
\vec{a}^{(SPH)}_i=-\sum_j m_j \left[\left({p_i\over\rho_i^2}+
    {p_j\over\rho_j^2}\right)+\Pi_{ij}\right]{\bf \nabla}_i W_{ij}.
    \label{fsph}
\end{equation}
Various forms for the AV term $\Pi_{ij}$ are discussed below (\S2.2.3).
The AV ensures that correct jump
conditions are satisfied across (smoothed) shock fronts, while the rest of
eq.~(\ref{fsph}) represents one of many possible SPH-estimators
for the acceleration due to the local pressure gradient (see, e.g.,
Monaghan 1985).

To provide reasonable accuracy, an SPH code must solve the equations of motion 
of a large number of particles (typically $N\gg 1000$). This rules out a
direct summation method for calculating the gravitational field of the
system, unless special purpose hardware such as the GRAPE is used
(Steinmetz 1996; Klessen 1997; see the article by Makino in this volume
for an update on GRAPE computers).  In most implementations of SPH,
particle-mesh algorithms (Evrard 1988; Rasio \& Shapiro
1992; Couchman et al.\ 1995) 
or tree-based algorithms (Hernquist \& Katz 1989; Dave et al.\ 1997) 
are used to calculate the gravitational accelerations $\vec{a}^{(Grav)}_i$.  
Tree-based algorithms perform better for problems involving large
dynamic ranges in density, such as star formation and large-scale cosmological
simulations. For typical stellar interaction problems,
density contrasts rarely exceed a factor $\sim10^2-10^3$ and 
grid-based algorithms and direct solvers are then generally faster.
Tree-based and grid-based algorithms can also be used to calculate lists
of nearest neighbors for each particle, exactly as in gravitational
$N$-body simulations.

\subsubsection{Artificial Viscosity}

For the AV, a symmetrized version of the form
proposed by Monaghan (1989) is often adopted,
\begin{equation}
\Pi_{ij}={-\alpha\mu_{ij}c_{ij}+\beta\mu_{ij}^2\over\rho_{ij}},
\label{pi}
\end{equation}
where $\alpha$ and $\beta$ are constant parameters,
$c_{ij}=(c_i+c_j)/2$ is the average sound speed, and
\begin{equation}
\mu_{ij}=\cases{ {\left(\vec{v}_i-\vec{v}_j\right)\cdot(\vec{r}_i-\vec{r}_j)\over
h_{ij}\left(|\vec{r}_i -\vec{r}_j|^2/h_{ij}^2+\eta^2\right)}& if $(\vec{v}_i-
\vec{v}_j)\cdot(\vec{r}_i-\vec{r}_j)<0$\cr
	     0& if $(\vec{v}_i-\vec{v}_j)\cdot(\vec{r}_i-
             \vec{r}_j)\ge0$\cr}
		\label{mu}
\end{equation}
with $h_{ij}=(h_i+h_j)/2$ and the constant $\eta^2\sim 10^{-2}$
(introduced to prevent numerical divergences).  
This form represents a combination of a bulk
viscosity (linear in $\mu_{ij}$) and a von~Neumann-Richtmyer 
viscosity (quadratic in $\mu_{ij}$) for convergent flows
($\mu_{ij}=0$ in regions of divergent flow).  The von~Neumann-Richtmyer
viscosity was initially introduced to suppress particle
interpenetration in the presence of strong shocks. For most problems,
eq.~(\ref{pi})
provides an optimal treatment of shocks when $\alpha\simeq 0.5$ and
$\beta\simeq 1$ (Monaghan 1989; Hernquist \&
Katz 1989; Lombardi et al.\ 1999).

A well-known problem with the classical AV of eq.~(\ref{pi})
is that it can generate
large amounts of spurious shear viscosity.  For this reason, Hernquist \& Katz (1989)
introduced another form for the AV:
\begin{equation}
\Pi_{ij}=\cases{ {q_i\over\rho_{i}^2}+{q_j\over\rho_{j}^2}& if $(
\vec{v}_i-\vec{v}_j)\cdot(\vec{r}_i-\vec{r}_j)<0$\cr
	     0& if $(\vec{v}_i-\vec{v}_j)\cdot(\vec{r}_i-
             \vec{r}_j)\ge0$\cr},
		\label{pi2}
\end{equation}
where
\begin{equation}
q_i=\cases{ \alpha \rho_i c_i h_i |{\bf \nabla}\cdot \vec{v}|_i+
	       \beta \rho_i h_i^2 |{\bf \nabla}\cdot \vec{v}|_i^2
		 & if $\left({\bf \nabla}\cdot \vec{v}\right)_i<0$\cr
             	0& if $\left({\bf \nabla}\cdot \vec{v}\right)_i\ge0$\cr}
			\label{q}
\end{equation}
and
\begin{equation}
({\bf \nabla}\cdot \vec{v})_i={1 \over \rho_i}\sum_j m_j
	(\vec{v}_j-\vec{v}_i)\cdot{\bf \nabla}_i W_{ij}. \label{divv}
\end{equation}
Although this form provides a slightly less accurate description of 
shocks than eq.~(\ref{pi}), it does
exhibit less shear viscosity.

More recently, Balsara (1995) has proposed the AV
\begin{equation}
\Pi_{ij}=
\left({p_i\over\rho_i^2}+{p_j\over\rho_j^2}\right)
	\left(-\alpha \mu_{ij} + \beta \mu_{ij}^2\right),
	\label{piDB}
\end{equation}
where
\begin{equation}
\mu_{ij}=\cases{ {(\vec{v}_i-\vec{v}_j)\cdot(\vec{r}_i-\vec{r}_j)\over
h_{ij}\left(|\vec{r}_i -\vec{r}_j|^2/h_{ij}^2+\eta^2\right)}{f_i+f_j \over 2
c_{ij}}& if $(\vec{v}_i-\vec{v}_j)\cdot(\vec{r}_i-\vec{r}_j)<0$\cr
	     0& if $(\vec{v}_i-\vec{v}_j)\cdot(\vec{r}_i-\vec{r}_j)\ge0$\cr}.
		\label{muDB}
\end{equation}
Here $f_i$ is the form function for particle $i$ defined by
\begin{equation}
f_i={|{\bf \nabla}\cdot \vec{v}|_i \over |{\bf \nabla}\cdot \vec{v}|_i
+|{\bf \nabla}\times \vec{v}|_i + \eta' c_i/h_i
}, \label{fi}
\end{equation}
where the factor $\eta'\sim 10^{-4}-10^{-5}$ prevents numerical
divergences, $({\bf \nabla}\cdot \vec{v})_i$ is given by eq.~(\ref{divv}), and
\begin{equation}
({\bf \nabla}\times \vec{v})_i={1 \over \rho_i}\sum_j m_j
	(\vec{v}_i-\vec{v}_j)\times{\bf \nabla}_i W_{ij}. \label{curlv}
\end{equation}
The form function $f_i$ acts as a switch, approaching unity
in regions of strong compression ($|{\bf \nabla}\cdot \vec{v}|_i
\gg |{\bf \nabla}\times \vec{v}|_i$) and vanishing in regions of large
vorticity ($|{\bf \nabla}\times \vec{v}|_i \gg |{\bf \nabla}\cdot 
\vec{v}|_i$).  Consequently, this AV has the advantage that 
dissipation in shear layers is suppressed.  
Note that since $(p_i/\rho_i^2+p_j/\rho_j^2)\simeq
2c_{ij}^2/(\gamma\rho_{ij})$, eq.~(\ref{piDB})
behaves like eq.~(\ref{pi}) when $|{\bf
\nabla}\cdot \vec{v}|_i \gg |{\bf \nabla}\times \vec{v}|_i$,
provided one rescales the $\alpha$ and $\beta$ in eq.~(\ref{piDB}) to be a
factor of $\gamma/2$ times the $\alpha$ and $\beta$ in eq.~(\ref{pi}).

For additional alternative treatments of AV, see the papers by Morris \& 
Monaghan (1997), Shapiro et al.\ (1996), Selhammar (1997), and Owen et al.\ (1998).

Note that a {\em physical viscosity\/} can also be added to SPH, for simulations
of 3D viscous flows, solving the Navier-Stokes equation
(e.g., Flebbe et al.\ 1994; Watkins et al.\ 1996). 

\subsubsection{Thermodynamics}

To complete the description of the fluid, either $u_i$ or $A_i$ is
evolved according to a discretized version of the first law of
thermodynamics.  Although various forms of these evolution equations exist, 
the most commonly used are
\begin{equation}
{ du_i\over dt}= {1\over 2}\sum_j m_j \left({p_i\over\rho_i^2}+{p_j\over\rho_j^2}+
    \Pi_{ij}\right)\,(\vec{v}_i-\vec{v}_j)\cdot{\bf \nabla}_i
    W_{ij},
	\label{udot}
\end{equation}
and
\begin{equation}
{dA_i\over dt}={\gamma-1\over 2\rho_i^{\gamma-1}}\,
     \sum_jm_j\,\Pi_{ij}\,\,(\vec{v}_i-\vec{v}_j)\cdot{\bf \nabla}_i
     W_{ij}.
	\label{adot}
\end{equation}
We call eq.~(\ref{udot}) the ``SPH energy equation,'' while eq.~(\ref{adot}) 
is the ``SPH entropy equation.''  Which equation one should
integrate depends upon the problem being treated.  Each has its own
advantages and disadvantages. Note that the simple forms of
eqs.~(\ref{udot}) and~(\ref{adot}) neglect terms proportional to
the time derivative of $h_i$.  Therefore, if we integrate the energy
equation, even in the absence of AV, the total
entropy of the system will not be strictly conserved if the particle
smoothing lengths are allowed to vary in time; if the entropy equation
is used to evolve the system, the total entropy is then strictly
conserved when $\Pi_{ij}=0$, but not the total energy (Rasio 1991;
Hernquist 1993). Both errors generally decrease as the number of
particles $N$ is increased (Rasio 1991), so that for large simulations
eqs.~(\ref{udot}) and~(\ref{adot}) are usually adequate.
For more accurate treatments involving
time-dependent smoothing lengths, see Nelson \& Papaloizou (1994)
and Serna et al.\ (1996).
The energy equation has the advantage that other thermodynamic processes
such as heating and cooling (Katz et al.\ 1996) and nuclear burning
(Garcia-Senz et al.\ 1998) can be incorporated more naturally.

\subsubsection{Integration in Time}

The results of SPH simulations involving only hydrodynamic forces 
and gravity do not depend strongly on the particular
time-stepping scheme used, as long as the timesteps are short enough
to maintain stability and
accuracy.  A simple second-order, explicit leap-frog scheme is often
employed. Implicit schemes must be used when other processes such as heating 
and cooling are coupled to the dynamics (Katz et al.\ 1996).
A low-order scheme is appropriate for SPH because pressure
gradient forces are subject to numerical noise.  For stability, the
timestep must satisfy a modified Courant condition, with $h_i$ replacing the
usual grid separation.
For accuracy, the timestep must be a small enough fraction of the
 dynamical time.

Among the many possible choices for determining the timestep,
the prescription proposed by Monaghan (1989) is recommended. This sets
\begin{equation}
\Delta t=C_N\,{\rm Min}(\Delta t_1,\Delta t_2), \label{dt}
\end{equation}
where the constant dimensionless Courant number $C_N$ typically
satisfies $0.1< C_N < 0.8$, and where
\begin{eqnarray}
\Delta t_1 &=&{\rm Min}_i\,(h_i/\dot v_i)^{1/2}, \label{dt1} \\
\Delta t_2 &=&{\rm Min}_i\left(
{h_i \over c_i+k\left(\alpha c_i+\beta {\rm Max_j}|\mu_{ij}|\right)}
\right),  \label{good.dt}
\end{eqnarray}
with $k$ being a constant of order unity.
If the Hernquist \& Katz AV [eq.~(\ref{pi2})] is used, the quantity
Max$_j|\mu_{ij}|$ in eq.~(\ref{good.dt}) can be replaced by $h_i|{\bf
\nabla}\cdot \vec{v}|_i$ if $({\bf \nabla}\cdot \vec{v})_i<0$, and by
$0$ otherwise.  By accounting for AV-induced diffusion, the $\alpha$ and 
$\beta$ terms in the denominator of eq.~(\ref{good.dt}) allow 
for a more efficient use of
computational resources than simply using a smaller value of $C_N$.

\subsubsection{Smoothing Lengths and Accuracy}

The size of the smoothing lengths is often chosen such that all particles
maintain approximately some predetermined number of neighbors $N_N$.  Typical
values of $N_N$ for 3D work range from about 20 to 100.  If a particle
interacts with too few neighbors, then the forces on it are sporadic, a
poor approximation to the forces on a true fluid element. In general, one
 finds that, for given physical conditions, 
the noise level in a calculation always decreases when $N_N$ is increased.  

At the other
extreme, large neighbor numbers degrade the resolution by requiring
unreasonably large smoothing lengths.  
However,  higher accuracy is obtained in SPH calculations only when 
{\em both\/}
the number of particles $N$ {\em and\/} the number of neighbors $N_N$ are
increased, with $N$ increasing faster than $N_N$ so that the smoothing
lengths $h_i$ decrease. Otherwise (e.g., if $N$ is increased while maintaining
$N_N$ constant) the SPH method is {\em inconsistent\/}, i.e., it converges to an
unphysical limit (see \S3 below). The choice of $N_N$ for a given
calculation is therefore dictated by a compromise between an acceptable
level of numerical noise and the desired spatial resolution (which is
$\simeq h\propto  N_N^{1/d}$ in $d$ dimensions) and level of accuracy.

\subsection{Test Calculations}

A number of numerical studies of the SPH scheme based on test calculations  
have been published recently.
A comprehensive study focusing on stellar interaction problems
was presented in Lombardi et al.\ (1999). 
In that study, a series of systematic tests were performed to evaluate
quantitatively the effects of spurious transport in 
3D SPH calculations. In particular, the tests
examined particle diffusion, numerical viscosity, and angular momentum 
transport. The main results can be summarized briefly as follows. 
Individual SPH particles behave like molecules in a liquid (or a solid for
the relaxed configurations often used as initial conditions).
Spurious particle diffusion in the ``liquid phase'' can be characterized by 
a diffusion coefficient $D$, which is a function of the ``temperature,''
or level of noise in the system, and the particle density $n$. For simulations 
with $N_N\simeq50-100$,
typical noise levels correspond to a Maxwellian distribution of
random particle velocities with $v_{rms}\simeq 0.05 c_s$, where $c_s$ is the
local (physical) speed of sound,
leading to a minimal amount of spurious diffusion,
with $D\simeq 0.005-0.02\,c_s n^{-1/3}$.  
A single set of values for the AV parameters $\alpha$ and
$\beta$ remain nearly optimal in a large number of situations: $\alpha\simeq 0.5$,
$\beta\simeq 1$ for the classical AV of Monaghan (eq.~[\ref{pi}]),
$\alpha\simeq\beta\simeq0.5$ for the Hernquist \& Katz AV (eq.~[\ref{pi2}]), and
$\alpha\simeq\beta\simeq\gamma/2$ for the Balsara AV (where $\gamma$
is the adiabatic index; see eq.~[\ref{piDB}]).  However these choices
may be modified depending on the goals of the particular
application.  For instance, if spurious particle mixing is not a
concern and only weak shocks (Mach number ${\cal M}\sim1-2$) are expected
during a calculation, then a smaller value of $\alpha$ is appropriate.
Somewhat larger values for $\alpha$ and $\beta$ may be preferable if an
accurate treatment of high Mach number shocks (${\cal M}\gg 1$) is
required.  Both the Hernquist \& Katz and Balsara forms
introduce relatively small amounts of numerical shear viscosity.
Furthermore, both Monaghan's and Balsara's AV do well at treating
shocks and at limiting the amount of spurious mixing.  

Other papers presenting results of more narrowly focused
sets of test calculations with SPH 
include those by Rasio \& Shapiro (1992), Navarro \& White (1993),
Steinmetz \& M\"uller (1993), Shapiro et al.\ (1996), and Owen et al.\ (1998). 
Comparisons between the results of SPH and Eulerian grid-based codes for the 
same problems have been
presented by Davies et al.\ (1993), Smith et al.\ (1996),
Bate \& Burkert (1997), and Sigalotti \& Klapp (1997).

\section{Convergence and Consistency}

For practical purposes, it is not enough to know that the
solution of the SPH equations will converge towards the exact fluid
solution of a problem in the limit where $N\rightarrow\infty$ and
$h\rightarrow0$. We are more interested in knowing how far from the
exact solution we can expect to be, for given computational resources. 
Moreover, since two independent parameters are involved (the total
number of particles $N$ and the smoothing length $h$), we would like to
know {\em how\/} this limit should be taken. If we wanted to
construct a hierarchy of increasingly accurate solutions, how should
$N$ and $h$ be changed from one calculation to the next? In this section, we
try to answer these  questions by studying a specific example in
detail.

The example we consider is the propagation of a linear sound wave in a
1D system. This example is simple enough that the SPH equations
can be solved {\em analytically\/}, allowing us to calculate the error
present in the SPH solution and to study its behavior as a function of 
$N$ and $h$, as well as other parameters.

The 1D gas is represented by an infinite string of 
SPH particles. All particles have the same mass $m$,
entropy constant $A$, and smoothing length $h$. The adiabatic SPH 
equations~(2.1), (2.6), and~(2.13) give
\begin{equation}
{d^2x_i\over dt^2}=-mA\,\sum_j\left(\rho_i^{\gamma-2}
         +\rho_j^{\gamma-2}\right)\, W'_{ij},
\end{equation}
where $W'_{ij}\equiv\partial W(x_i-x_j)/\partial x_i$.
If the kernel $W(x)$ is an even function, which we assume, it is clear
that the solution $x_i=ia$, ${\dot x}_i=0$, for $i=-\infty,+\infty$,
describes an equilibrium. Here $a$ is the interparticle separation.
In practice the system would be represented by a segment of finite
length $L$ with periodic boundary conditions, so that $a=L/N$ is just a
measure of the total number of particles. Throughout the calculation,
however, we assume that $L\gg h$ and $L\gg \lambda$, where $\lambda$ is the
wavelength, so that the effects of the boundary conditions can be ignored.

We now perturb the equilibrium, writing 
\begin{equation}
x_i=ia+\delta x_i,
\end{equation}
 and
\begin{equation}
\rho_i=\rho_0+\delta\rho_i,
\end{equation}
where $\rho_0=m/a$ is the equilibrium density (in 1D).
We expand the density as 
\begin{equation}
\rho_i^{\gamma-2}=\rho_0^{\gamma-2}
        +(\gamma-2)\rho_0^{\gamma-3}\delta\rho_i+\ldots,
\end{equation}
and the kernel as
\begin{eqnarray}
    W_{ij}=W\bigl((i-j)a\bigr)
         &+&(\delta x_i-\delta x_j)W'\bigl((i-j)a\bigr) \nonumber \\
         &+&{1\over2}(\delta x_i-\delta x_j)^2
               W''\bigl((i-j)a\bigr)+\ldots,
\end{eqnarray}
and
\begin{equation}
W'_{ij}=W'\bigl((i-j)a\bigr)
      +(\delta x_i-\delta x_j)W''\bigl((i-j)a\bigr)+\ldots.
\end{equation}
We then linearize the equations of motion (3.1) and find, after some algebra,
\begin{eqnarray}
   {d^2\delta x_i\over dt^2} = -m&A&\rho_0^{\gamma-2}\Bigl[\,2\sum_j
  (\delta x_i-\delta x_j)W''\bigl((i-j)a\bigr) \nonumber \\
  &+&{(\gamma-2)\delta\rho_i\over\rho_0}\sum_jW'\bigl((i-j)a\bigr) \nonumber \\
  &+&{(\gamma-2)\over\rho_0}\sum_j\delta\rho_j
       W'\bigl((i-j)a\bigr)\,\Bigr].
\end{eqnarray}
The second sum on the right vanishes identically for an even kernel.
We now let $\delta x_i\propto \exp[{\rm i}\,(kia-\omega t)]$. The first 
sum in the right hand side of eq.~(3.7) becomes
\begin{equation}
2\sum_j(\delta x_i-\delta x_j)W''\bigl((i-j)a\bigr)=
2\, \delta x_i \sum_n\bigl[1-\exp(-{\rm i}\,kna)\bigr]W''\bigl(na\bigr),
\end{equation}
where we have defined a new index $n=i-j$.
Similarly, using eqs.~(2.5) and~(3.5) for the density, the third sum becomes
\begin{equation}
(\gamma-2)\frac{\delta x_i}{\rho_0}\, m\sum_n\sum_m\exp(-{\rm i}\,kna)
\bigl[1-\exp(-{\rm i}\,kma)\bigr]W'(na)W'(ma).
\end{equation}
Combining eqs.~(3.7)---(3.9) we find the {\em dispersion relation\/},
\begin{eqnarray}
  & \omega^2 &= m A\rho_0^{\gamma-2}\,\Bigl[2\sum_n\bigl[1-\exp(-{\rm i}\,
  kna)\bigr]W''(na) \nonumber \\
   &+& (\gamma-2)a\sum_n\sum_m
  \exp(-{\rm i}\,kna)\bigl[1-\exp(-{\rm i}\,kma)\bigr]
         W'(na)W'(ma)\,\Bigr].
\end{eqnarray}
We are interested in taking two different limits of this equation,
and comparing them to the exact dispersion relation, $\omega^2
=k^2c_0^2$, where $c_0=(\gamma p_0/\rho_0)^{1/2}=$ $(\gamma
A\rho_0^{\gamma-1})^{1/2}$ is the speed of sound. 

The first limit of interest is the long wavelength limit, which
corresponds to $a/\lambda\rightarrow0$. This is equivalent to 
the limit of a {\em large number of particles\/},
$N\rightarrow\infty$.  In taking this limit, we do not assume anything
about the ratio $h/a$, so that the number of neighbors $N_N$ remains
arbitrary. Taking the $k\rightarrow0$ limit of eq.~(3.10) and 
defining $c^2_{SPH}\equiv \omega^2/k^2$ we get
\begin{equation}
{c_{SPH}^2\over c_0^2}={2\over\gamma}\bigl({a\over
   h}\bigr)^3\sum_{n>0}n^2w''\bigl[n\bigl({a\over h}
   \bigr)\bigr]+4\,{\gamma-2\over\gamma}\bigl({a\over
   h}\bigr)^4\,\left(\sum_{n>0}n\,w'\bigl[n\bigl({a\over h}\bigr)
     \bigr]\right)^2.
\end{equation}
The three relevant parameters in this expression are the dimensionless form
of the kernel, $w(\xi)\equiv h W(h\xi)$, the ratio $h/a\propto N_N$, and the
adiabatic exponent $\gamma$. It is easy to see from eq.~(3.11) that 
$w(\xi)$ should be twice differentiable and that the correct 
physical result is then recovered only in the limit where $N_N\rightarrow\infty$. 
One implication, which is of great practical importance, is that the combined
limit $N\rightarrow\infty$ and $h\rightarrow0$ should not be constructed
by letting the number of particles increase while keeping
$N_N$ constant. Taking the limit
in this way results in an {\em inconsistent scheme\/}, one which does
converge, but not towards the correct result. Instead, both $N_N$ and $N$
should be increased, with $N$ increasing faster than $N_N$ so that
$h\rightarrow0$ in the process. 
In addition, it is easy to show (e.g., by evaluating eq.~[3.11] for different
spline kernels of increasing order) that the convergence is
faster when a smoother kernel is used. Therefore,
for given computational resources, increasing the
smoothness of the kernel can improve the accuracy of the results.
Note also that the convergence is faster for larger values of the
adiabatic exponent $\gamma$, i.e., for more incompressible fluids.

The other limit of interest for eq.~(3.10) is when the ratio
$h/a\rightarrow\infty$, which corresponds to the limit of a {\em large number
of neighbors\/}, $N_N\rightarrow\infty$. In this limit, all the sums in
eq.~(3.10) can be replaced by integrals, using the substitution 
$\sum_n(\ldots)\,\rightarrow\,\int (\ldots)dx/a$, with $x=na$.
After some algebra we find,
\begin{equation}
{c_{SPH}^2\over c_0^2}={2\over\gamma}\int_{-\infty}^{+\infty}
    \cos kx\,W(x)dx+(1-{2\over\gamma})\bigl(\int_{-\infty}^{+\infty}
        \cos kx\,W(x)dx\bigr)^2.
\end{equation}
Note that this expression is valid for any $k$, since we did not
take the long wavelength limit.

Finally, we can combine both limits by taking the $k\rightarrow0$
limit of eq.~(3.12), which gives
\begin{equation}
{c_{SPH}^2\over c_0^2}=1-{\gamma-1\over\gamma}
   k^2\int_{-\infty}^{+\infty}
          x^2W(x)dx+{\cal O}(k^4).
\end{equation}
This equation demonstrates explicitly the consistency of the scheme
in the limit  ($N$, $N_N$) $\rightarrow\infty$ and $h\rightarrow0$.
The leading term of the error is, in general, ${\cal O}(k^2)$. The
only exceptions are when $\gamma=1$ (the isothermal case)
\footnote{Monaghan (1989) performed a similar calculation but
considered only, for simplicity, the isothermal case. This led him
to conclude, incorrectly, that the leading error term in the dispersion
relation was in general ${\cal O}(k^6)$ (cf.~his eq.~[3.11]).},
and when $W(x)$ is such that the integral in eq.~(3.13) 
vanishes. This would require the kernel to be nonpositive. An example 
of a 1D kernel having this property is the $W_4$ kernel
of Monaghan (1985):
\begin{equation}
W_4(x;h)={1\over h}\cases{
   1-{5\over2}\left(\frac{x}{h}\right)^2+{3\over2}\left(\frac{x}{h}\right)^3,   
           & $0\le \frac{x}{h}<1$;\cr
   {1\over2}(1-\frac{x}{h})(2-\frac{x}{h})^2,  & $1\le \frac{x}{h}<2$;\cr
    0,  &otherwise.\cr} 
\end{equation}
This kernel is negative over the interval $1< x/h <2$ and has a continuous
first derivative. Naturally, one should be careful when using such a
kernel in SPH, since physical quantities such as pressure and density
are no longer guaranteed to remain positive.

We conclude by summarizing some of our key results:

\smallskip\noindent
(a) Higher accuracy is obtained in SPH calculations when {\em both\/}
the number of particles $N$ {\em and\/} the number of neighbors $N_N$ are
increased, with $N$ increasing faster than $N_N$ so that the smoothing
length $h$ decreases. 

\smallskip\noindent
(b) A number of neighbors $N_N\gg 1$ must be
maintained at all times for a calculation to be meaningful.

\smallskip\noindent
(c) The SPH scheme is consistent in the limit where
$N\rightarrow\infty$, $N_N\rightarrow\infty$, and $h\rightarrow0$.

\smallskip\noindent
(d) Convergence can be accelerated significantly by increasing the
smoothness of the kernel.

\smallskip\noindent
These results were obtained here on the basis of a particularly
simple example, but we expect them to be applicable in general.
In particular, it is easy to apply this analysis to other forms
of the SPH scheme (e.g., using eq.~[\ref{XSPH}] instead of eq.~[\ref{rdot}]), 
or to extend it to
more than one dimension, which allows a study of spurious anisotropy
effects (Kalogera \& Rasio 1999).

\section*{Acknowledgements}
This work was supported by NSF Grant AST-9618116, NASA ATP
Grant NAG5-8460, and a Sloan Research Fellowship. 
Our computational work was supported by the National 
Computational Science Alliance and utilized the SGI/Cray Origin2000
supercomputer at NCSA.


\newpage

\begin{thebibliography}{99}
\bibitem{} Bailey, V.C., \& Davies, M.B. 1999, MNRAS, 308, 257
\bibitem{} Balsara, D. 1995, J. Comp. Phys., 121, 357
\bibitem{} Bate, M.R., \& Burkert, A. 1997, MNRAS, 288, 1060
\bibitem{} Burkert, A., Bate, M.R., \& Bodenheimer, P. 1997, MNRAS, 289, 497
\bibitem{} Couchman, H.M.P., Thomas, P.A., \& Pearce, F.R. 1995, ApJ, 452, 797
\bibitem{} Dave, R., Dubinski, J., \& Hernquist, L. 1997, New A, 2, 277
\bibitem{} Davies, M.B., Ruffert, M., Benz, W., Muller, E. 1993, A\&A, 272, 430
\bibitem{} Evrard, A.E. 1988, MNRAS, 235, 911
\bibitem{} Flebbe, O., Muenzel, S., Herold, H., Riffert, H., Ruder, H. 1994, ApJ, 431, 754
\bibitem{} Fulbright, M.S., Benz, W., \& Davies, M.B. 1995, ApJ, 440, 254
\bibitem{} Garcia-Senz, D., Bravo, E., \& Serichol, N. 1998, ApJS, 115, 119
\bibitem{} Gingold, R.A., \& Monaghan, J.J. 1977, ApJ, 181, 375
\bibitem{} Gingold, R.A., \& Monaghan, J.J. 1982, J.\ Comp.\ Phys., 46, 429
\bibitem{} Herant, M., \& Benz, W. 1992, ApJ, 387, 294
\bibitem{} Hernquist, L. 1993, ApJ, 404, 717
\bibitem{} Hernquist, L., \& Katz, N. 1989, ApJS, 70, 419
\bibitem{} Kalogera, V., \& Rasio, F.A. 1999, in preparation
\bibitem{} Katz, N. 1992, ApJ, 391, 502
\bibitem{} Katz, N., Weinberg, D.H., \& Hernquist, L. 1996, ApJS, 105, 19
\bibitem{} Klessen, R. 1997, MNRAS, 292, 11
\bibitem{} Lombardi, J.C., Jr., Rasio, F.A., \& Shapiro, S.L. 1995, ApJ, 445, L117
\bibitem{} Lombardi, J.C., Jr., Rasio, F.A., \& Shapiro, S.L. 1996, ApJ, 468, 797
\bibitem{} Lombardi, J.C., Jr., Rasio, F.A., Sills, A. \& Shapiro, S.L. 1999, 
   J.\ Comp.\ Phys., 152, 687
\bibitem{} Lucy, L. B. 1977, AJ, 82, 1013
\bibitem{} Monaghan, J.J.\ 1985, Comp. Phys. Rep., 3, 71
\bibitem{} Monaghan, J.J.\ 1989, J. Comp. Phys., 82, 1
\bibitem{} Monaghan, J.J.\ 1992, ARA\&A, 30, 543
\bibitem{} Monaghan, J.J., \& Lattanzio, J.C. 1985, A\&A, 149, 135
\bibitem{} Morris, J.P., \& Monaghan, J.J. 1997, J.\ Comp.\ Phys., 136, 41
\bibitem{} Navarro, J.F., \& White, S.D.M. 1993, MNRAS, 265, 271
\bibitem{} Nelson, A.F., Benz, W., Adams F.C., \& Arnett, D. 1998, ApJ, 502, 342
\bibitem{} Nelson, R.P., \& Papaloizou, J.C.B. 1994, MNRAS, 270, 1
\bibitem{} Owen, J.M., Villumsen, J.V., Shapiro, P.R., Martel, H. 1998, ApJS, 116, 155O
\bibitem{} Rasio, F.A. 1991, PhD Thesis, Cornell University
\bibitem{} Rasio, F.A., \& Livio, M. 1996, ApJ, 471, 366 
\bibitem{} Rasio, F.A., \& Shapiro, S.L. 1992, ApJ, 401, 226
\bibitem{} Rasio, F.A., \& Shapiro, S.L. 1994, ApJ, 432, 242
\bibitem{} Rasio, F.A., \& Shapiro, S.L. 1995, ApJ, 438, 887
\bibitem{} Rosswog, S., Liebendoerfer, M., Thielemann, F.-K., Davies, M.B., Benz, W.,
  \& Piran, T. 1999, A\&A, in press 
\bibitem{} Ruffert, M. 1993, A\&A, 280, 141
\bibitem{} Selhammar, M. 1997, A\&A, 325, 857
\bibitem{} Serna, A., Alimi, J.-M.,  \& Chi\`eze, J.-P. 1996, ApJ, 461, 884
\bibitem{} Shapiro, P.R., Martel, H., Villumsen, J.V., Owen, J.M. 1996, ApJS, 103, 269
\bibitem{} Sigalotti, L., \& Klapp, J. 1997, A\&A, 319, 547
\bibitem{} Smith, S.C., Houser, J.L., \& Centrella, J.M. 1996, ApJ, 458, 236
\bibitem{} Steinmetz, M. 1996, MNRAS, 278, 1005
\bibitem{} Steinmetz, M., \& M\"uller, E. 1993, A\&A, 268, 391
\bibitem{} Watkins, S.J., Bhattal, A.S., Francis, N., Turner, J.A., 
  \& Whitworth, A.P. 1996, A\&AS, 119, 177
\bibitem{} Woodward, P.R. 1986, in  Astrophysical Radiation Hydrodynamics,
  eds.\ K.A.~Winkler \& M.L.~Norman (NATO ASI Series, Dordrecht: Reidel) 245
\bibitem{} Zhuge, X., Centrella, J.M., \& McMillan, S.L.W. 1996, 
   PRD, 54, 7261
\end{thebibliography}
\end{document}